\documentclass[twocolumn,showpacs,floatfix,epsf,psfig]{revtex4}

\usepackage{epsf}
\usepackage{psfig}
\usepackage{graphicx}
\begin{document}

\title
{Magnetic  properties and electronic structure of Mn-Ni-Ga magnetic shape memory alloys}

\author{Sunil Wilfred D$^{\prime}$Souza$^1$, Tufan Roy$^2$,  Sudipta Roy Barman$^{1}$ and Aparna Chakrabarti$^2$} 

\affiliation{$^1$UGC-DAE Consortium for Scientific Research, Khandwa Road, Indore, 452001, Madhya Pradesh, India}
\affiliation{$^2$Raja Ramanna Centre for Advanced Technology, Indore, 452013, Madhya Pradesh, India}

\begin{abstract}
Influence of disorder, antisite defects, martensite transition and compositional variation on the magnetic properties and electronic structure of Mn$_2$NiGa and  Mn$_{1+x}$Ni$_{2-x}$Ga magnetic shape memory alloys have been studied by using  full potential spin-polarized scalar relativistic Korringa-Kohn-Rostocker (FP-SPRKKR) method.  
Mn$_2$NiGa is ferrimagnetic and its total spin moment  increases when disorder in the occupancy of Mn$_{\rm Ni}$ (Mn atom in Ni position) is considered. The moment further increases when    Mn-Ga antisite defect\cite{Singh12} is included in the calculation.  A reasonable estimate of $T_C$ for Mn$_2$NiGa is obtained from the exchange parameters for the disordered structure. Disorder influences the electronic structure of Mn$_2$NiGa through overall broadening of the density of states and a decrease in the exchange splitting. Inclusion of antisite defects marginally broaden the minority spin partial DOS (PDOS), while the majority spin PDOS is hardly affected. For Mn$_{1+x}$Ni$_{2-x}$Ga where 1$\geq$$x$$\geq$0, as $x$ decreases, Mn$_{\rm Mn}$ moment increases while Mn$_{\rm Ni}$ moment decreases in both austenite and martensite phases. For $x$$\geq$~0.25, the total moment of the martensite phase is smaller compared to the austenite phase, which indicates possible occurrence of inverse magnetocaloric effect. We find that the redistribution of Ni 3$d$- Mn$_{\rm Ni}$ 3$d$ minority spin  electron states close to the Fermi level is primarily responsible for the stability of the martensite phase in Mn-Ni-Ga. 
\end{abstract}

\maketitle

\section {Introduction}
Mn$_2$NiGa is a magnetic shape memory alloy (MSMA) that has become  focus of intensive research due to its important properties of technological relevance such as sizable magnetic field induced strain of about 4$\%$,  high Curie temperature (588K),\cite{Liu05} and recently discovered spin-valve like magnetoresistance behavior.\cite{Singh12}  However, there are only a few theoretical studies in the literature that provide a basic understanding of the electronic structure and magnetic properties in this system. {\it Ab initio} density functional theory based calculation using full potential linearized augmented plane wave (FPLAPW) method has shown that Mn$_2$NiGa can undergo a volume conserving structural transistion from austenite to martensite phase, with $c/a$= 1.25 implying that this material can exhibit shape memory effect.\cite{Barman07} It was demonstrated by considering different starting magnetic configurations that  the  ground state of this material is ferrimagnetic with antiparallel coupling between the two nearest neighbor Mn atoms in both the austenite and martensite phases.\cite{Barman08} The ferrimagnetic state originates from the difference in hybridization of the majority and minority spin states: the hybridization
between the majority spin Ni and Mn$_{\rm Mn}$ 3$d$ states is stronger than the hybridization between Ni and Mn$_{\rm Ni}$  3$d$ minority spin states (the notations such as Mn$_{\rm Mn}$ and Mn$_{\rm Ni}$ are defined in the caption of Table I).\cite{Barman07} It was pointed out that the self consistent calculations could converge to a local minimum that is not the actual ground state,\cite{Barman08} as was the case in Ref.\cite{Liu06}  where, in the martensite phase, Mn$_2$NiGa was claimed to  be ferromagnetic (FM)  with almost zero moment on Mn(B) atoms.  
~Subsequently, different theoretical studies on related  MSMA systems such as Mn$_2$NiIn,\cite{Chakrabarti09}  Mn$_2$NiAl,\cite{Luo10} 
~and Mn$_2$NiSn\cite{Paul13}  have reported antiparallel coupling of the Mn$_{\rm Mn}$ and Mn$_{\rm Ni}$ moments.  
Using total energy minimization and considering different starting magnetic states, Mn$_2$NiIn was predicted to be a MSMA with ferrimagnetic ground state and stable tetragonal structure with $c/a$=~0.967.\cite{Chakrabarti09} A martensite
transformation has indeed been observed experimentally in Mn$_2$Ni$_{1.6}$In$_{0.4}$ ribbons.\cite{Llamazares08} 
~It has been shown that existence of disorder between Mn and Ni atoms in Mn$_2$NiSn  increases the magnetic moment compared to the  ordered structure.\cite{Paul13}    

Powder diffraction studies  reported that  Mn$_2$NiGa has a disordered cubic (L2$_1$) structure in the austenite phase with $Fm\bar{3}m$ space group,\cite{Brown10,Singh10} in disagreement with an earlier x-ray diffraction (XRD) study that showed that Mn$_2$NiGa has ordered Hg$_2$CuTi type inverse Heusler structure with $F\bar{4}3m$ space group.\cite{Liu06} XRD studies showed that  the structure of Mn$_2$NiGa is highly dependent on residual stress.\cite{Singh10} 
~The total moment in the martensite phase was found to be lower than  the austenite phase by theory,\cite{Barman07} which was later confirmed by magnetization and Compton scattering studies.\cite{Ahuja09} 

Recently, an unusual asymmetric magnetoresistance variation indicating a spin-valve like behavior in Mn$_2$NiGa was attributed to the formation of ferromagnetic clusters in the ferrimagnetic lattice.\cite{Singh12} This was concluded on the basis of neutron diffraction studies that  established existence of about 13\% Mn-Ga antisite defects in Mn$_2$NiGa lattice that is ferrimagnetic.  It was established by our spin polarized relativistic Korringa$-$Kohn$-$Rostoker (SPRKKR) calculations\cite{Singh12} that  ferromagnetic clusters occur because of the Mn-Ga antisite defects. Thus, disorder and defects play important role in Mn$_2$NiGa, and it is crucial to understand their influence on its electronic structure and magnetic properties.  So, in this paper, using  the SPRKKR method, we investigate these issues  in  Mn-Ni-Ga  (Mn$_{1+x}$Ni$_{2-x}$Ga where 1$\geq$$x$$\geq$0) with major emphasis on Mn$_2$NiGa. The paper is organized as follows: The Computational Details section discusses the method of calculation and provides the details of the structures used. The  Results and Discussions section is divided into six sub-sections: In the first two sections (A and B), the behavior of the magnetic moments of Mn-Ni-Ga are discussed. Subsequently, the exchange interaction parameters and the Curie temperature of Mn$_2$NiGa are discussed (section C).  The next two sub-sections deal with the effect of disorder and martensite transition on the electronic structure of Mn$_2$NiGa. Section F discusses the influence of compositional variation on the electronic structure.

\begin{table*}[htpb]  
\caption{The crystal structure, atomic positions and occupancies in the austenite and martensite phases of Mn-Ni-Ga. Note that Mn$_{\rm X}$ refers to a Mn atom at the X (X= Ni, Mn, Ga) atom site of Ni$_2$MnGa. In the austenite phase, 
~4$a$, 4$b$ and 8$c$ sites are occupied by Ga, Mn and Ni, respectively.}
\centering
\begin{tabular}{|c|c|c|c|c|} \hline
 \multicolumn{5} {|c|}  {{\bf Austenite}} \\\hline
 \multicolumn{5} {|c|}  {space group: $Fm$$\bar{3}$$m$, $c/a$= 1, L2$_1$ structure} \\\hline
  sites & 4a & 4b & 8c &\\\hline
 {\bf D$_{\rm A}$} (1$\geq$$x$$\geq$0) & Ga & Mn$_{\rm Mn}$ & (1-$x$)Ni + $x$Mn$_{\rm Ni}$ & \\\hline
  {\bf DwA$_{\rm A}$} ($x$=~1) & 0.88Ga + 0.12Mn$_{\rm Ga}$ & Mn$_{\rm Mn}$ & 0.5Ni + 0.44Mn$_{\rm Ni}$ + 0.06Ga$_{\rm Ni}$ & \\\hline
 \multicolumn{5} {|c|}  {space group: $F$$\bar{4}$$3m$, $c/a$= 1, inverse Heusler  Hg$_2$CuTi structure} \\\hline 
   sites & 4a & 4b & 4c & 4d\\\hline 
  {\bf O$_{\rm A}$} ($x$= 1) & Ga & Mn$_{\rm Mn}$ & Ni & Mn$_{\rm Ni}$ \\\hline \hline
   \multicolumn{5} {|c|}  {{\bf Martensite}} \\\hline
    \multicolumn{5} {|c|}  {space group: $Fmmm$, c/a= 1.25} \\\hline
      sites & 4a & 4b & 8f & \\\hline
  {\bf D$_{\rm M}$} (1$\geq$$x$$\geq$0)  & Ga & Mn$_{\rm Mn}$ & (1-$x$)Ni + $x$Mn$_{\rm Ni}$ & \\\hline
            \multicolumn{5} {|c|}  {space group: $I4/mmm$} \\\hline
                  sites & 2a & 2b & 4d & \\\hline
 {\bf DwA$_{\rm M}$} ($x$=~1) & Mn$_{\rm Mn}$ & 0.88Ga + 0.12Mn$_{\rm Ga}$ & 0.5Ni+0.44Mn$_{\rm Ni}$+0.06Ga$_{\rm Ni}$ & \\\hline
\end{tabular}
\end{table*}



\section {Computational Details}
The electronic structure calculations have been carried out by full potential spin polarized scalar relativistic Korringa-Kohn-Rostocker  (FP-SPRKKR) Green's function technique.\cite{Ebert} Substitutional disorder has been accounted for by the coherent potential approximation (CPA). The exchange-correlation effects are taken into consideration by using the generalized gradient approximation (GGA) in the Perdew-Burke-Ernzerhof parametrization scheme.\cite{Perdew} An angular momentum expansion up to $l_{max}$=~3 on a 22$\times$~22$\times$~22 $k$-point mesh in the irreducible wedge of the Brillouin zone was used to obtain an accurate ground state potential through the self-consistent cycles. Both the energy convergence criterion and the CPA tolerance were set to 10$^{-5}$ Ry. 

Table~I depicts the different crystal structures that have been used in this work. In the austenite phase, disorder has been considered by using the $Fm$$\overline{3}$$m$ space group and this structure is  henceforth referred to in the text as {\bf D$_{\rm A}$} ({\bf D}isordered {\bf A}ustenite). For the purpose of comparison, the calculations have also been performed using the ordered inverse Heusler structure referred to as {\bf O$_{\rm A}$} ({\bf O}rdered {\bf A}ustenite). Disorder in the martensite phase has been considered through the $Fmmm$ space group and is referred to as {\bf D$_{\rm M}$} ({\bf D}isordered {\bf M}artensite).  We have also carried out the SPRKKR calculation for Mn$_2$NiGa using the recent structure provided by neutron diffraction\cite{Singh12} that reports presence of Mn-\,Ga antisite defects  for both the {\bf A}ustenite (referred to as {\bf DwA$_{\rm A}$}  {\it i.e.} {\bf D}isordered {\bf w}ith {\bf A}ntisite defect) and {\bf M}artensite ({\bf DwA$_{\rm M}$}) phases to examine the effect of the antisite defects on the magnetic properties and the electronic structure (Table I). In the above notations, if the subscript ({\bf A} and {\bf M} indicating austenite and  martensite phase, respectively) is not used,  both the phases are indicated.

\begin{table*}  
\caption{Spin magnetic moments ($\mu_{B}$/f.u.) of Mn$_2$NiGa-{\bf O$_{\rm A}$} and Ni$_2$MnGa in the austenite phase.}
\centering
\begin{tabular}{|c|c|c|c|c||c|c|c|c||c|c|c|c|} \hline
 {composition} & \multicolumn{4} {|c||} {SPRKKR (ASA)} & \multicolumn{4} {|c||} {FP-SPRKKR} & \multicolumn{4} {|c|}  {FPLAPW}    \\ \hline
  & Mn$_{\rm Mn}$ & Ni & Mn$_{\rm Ni}$ & total & Mn$_{\rm Mn}$ & Ni & Mn$_{\rm Ni}$ & total & Mn$_{\rm Mn}$ & Ni & Mn$_{\rm Ni}$ & total    \\ 
    &  &  &  & spin &  &  &  & spin &  &  &  & spin     \\ \hline
  Mn$_{2}$NiGa-{\bf O$_{\rm A}$}  & 3.42 & 0.24 & -2.76 & 0.9 & 3.31 & 0.34 & -2.52 & 1.13 & 3.2 & 0.32 & -2.43 & 1.14 \\ \hline                
  Ni$_{2}$MnGa  & 3.54 & 0.32 & - &4.09 & 3.52 & 0.34 & -  & 4.14  &3.44 & 0.36 & -   & 4.13 \\ \hline                \end{tabular}
\vskip 1cm

\caption{Spin magnetic moments ($\mu_{B}$/f.u.) of the disordered and ordered structures of Mn$_{2}$NiGa in austenite and martensite phases using FP-SPRKKR method. }
\centering
\begin{tabular}{|c|c|c|c|c|c||c|c|c|c|c|c|} \hline
 {Mn$_{2}$NiGa} & \multicolumn{5} {|c||} {austenite} & {Mn$_{2}$NiGa} & \multicolumn{5} {|c|}  {martensite}    \\ 
  {structures} & \multicolumn{5} {|c||} { } & { structures} & \multicolumn{5} {|c|} {  }  \\ \hline
  & Mn$_{\rm Mn}$ & Ni & Mn$_{\rm Ni}$ & Mn$_{\rm Ga}$ & total &   & Mn$_{\rm Mn}$ & Ni & Mn$_{\rm Ni}$ & Mn$_{\rm Ga}$ & total      \\ \hline        
{\bf O$_{\rm A}$} & 3.31 & 0.34 & -2.52 & - & 1.13 &  &  &  &  &  &    \\ \hline               
{\bf D$_{\rm A}$}  & 3.29 & 0.43 & -1.5 & - & 2.22 & {\bf D$_{\rm M}$} & 3.12 & 0.34 & -2.21 & - & 1.25  \\ \hline
{\bf DwA$_{\rm A}$}  & 3.28 & 0.46 & -1.47 & 3.38 & 2.85 & {\bf DwA$_{\rm M}$} & 3.29 & 0.42 & -2.32 & 3.39 & 2.07 \\ \hline
\end{tabular}
\end{table*}

\vskip 1cm
\begin{table}  
\caption{ The starting and converged Mn$_{\rm Mn}$, Mn$_{\rm Ga}$ and Mn$_{\rm Ni}$ spin magnetic moments ($\mu_B$/f.u.) of Mn$_2$NiGa-{\bf DwA$_{\rm M}$} along with the corresponding converged total energies ($E_{tot}$). The lowest $E_{tot}$ is taken to be zero meV.} 
\centering
\begin{tabular}{|p{1cm}|p{1cm}|p{1cm}|p{1cm}|p{1cm}|p{1cm}|p{1cm}|} \hline
 \multicolumn{3} {|p{3cm}|} {Starting  moments } & \multicolumn{3} {|p{3cm}|}  {Converged moments} & {$E_{tot}$ (meV)}   \\\hline
  Mn$_{\rm Mn}$ & Mn$_{\rm Ga}$ & Mn$_{\rm Ni}$ & Mn$_{\rm Mn}$ & Mn$_{\rm Ga}$ & Mn$_{\rm Ni}$ &  \\ \hline 
  3.00 & 3.00 & -3.00  & 3.29 & 3.39 & -2.32  & 0 \\ \hline 
  3.00 & -3.00 & -3.00 & 3.27 & -3.29 & -2.38 & 48 \\ \hline 
  3.00  & 3.00 & 3.00  & 2.98  & 3.14 & 2.25  & 381 \\ \hline 
\end{tabular}
\end{table}

\section{Results and Discussion}

\subsection {Magnetic moments of Mn$_2$NiGa}
The magnetic moments of Mn$_2$NiGa-{\bf O$_{\rm A}$} in Table~II 
~portrays a ferrimagnetic ground state where Mn$_{\rm Mn}$ and Mn$_{\rm Ni}$ spin moments are antiparallel.  Calculations with Mn$_{\rm Ni}$ parallel to Mn$_{\rm Mn}$ resulted in 638 meV larger total energy. The results of atomic sphere approximation (ASA) and FP-SPRKKR calculations are similar, although  Mn$_{\rm Ni}$ moment is somewhat overestimated in ASA, resulting in marginally smaller total spin moment (Table~II). The individual orbital moments calculated by fully relativistic ASA SPRKKR method shows that these values are negligible compared to the spin moments, and hence can be neglected.\cite{Sunilcondmat}  Table~II also shows that the moments obtained from FP-SPRKKR  for both Mn$_2$NiGa-{\bf O$_{\rm A}$} and Ni$_2$MnGa are in excellent agreement with the moments calculated by the FPLAPW method.\cite{Barman07,Chakrabarti09,Ayuela02,Barman05}

In Mn$_2$NiGa-{\bf O$_{\rm A}$}, which is the ordered inverse Heusler structure, Mn$_{\rm Ni}$ occupies one of the two allowed Ni positions: (0.25, 0.25, 0.25) or (0.75, 0.75, 0.75). On the other hand in  Mn$_2$NiGa-{\bf D$_{\rm A}$}, Mn gets randomly distributed in both the Ni sites. Interesting influence of disorder on the magnetic moments  is evident from Table~III.  The total spin moment in the austenite phase increases from 1.13 to 2.22~$\mu_B$/f.u.  between {\bf O$_{\rm A}$} and {\bf D$_{\rm A}$}. This increase is due to the decrease in the magnitude of the antiparallel Mn$_{\rm Ni}$ moment  from 2.52 to 1.5 $\mu_B$/f.u. in presence of disorder, 
~while the Mn$_{\rm Mn}$ and Ni parallel moments remain largely unchanged. The origin of this effect can be traced to substantial changes in the  density of states (discussed latter in Section C). 

In Mn$_2$NiGa-{\bf DwA}, where there are 
~multiple magnetic interactions, the self-consistency calculations might converge to a local minimum depending on the starting magnetic moment configuration.\cite{Barman07,Barman08,Chakrabarti09} So, to obtain the lowest energy magnetic state, different starting configurations of Mn$_{\rm Ni}$, Mn$_{\rm Mn}$ and Mn$_{\rm Ga}$ moments have been considered (Table IV).\cite{Singh12}   We find that for the ferromagnetic state, where all the three types of Mn atoms are parallel, the total energy ($E_{tot}$) is largest (381 meV, third row in Table IV). The ground state is obtained when Mn$_{\rm Ga}$  is parallel  to Mn$_{\rm Mn}$, but anti-parallel to Mn$_{\rm Ni}$ spin moments. Thus, SPRKKR theory supports the results obtained from neutron diffraction.\cite{Singh12} 
~While the Mn$_{\rm Ni}$ and Mn$_{\rm Mn}$ are antiparallel because of direct interaction  at relatively short nearest neighbor distances,\cite{Barman08,Chakrabarti09,Hobbs03}  Mn$_{\rm Ga}$ being the next nearest neighbor of Mn$_{\rm Mn}$ at a larger separation, their interaction is ferromagnetic. 

Having established the magnetic ground state of Mn$_2$NiGa-{\bf DwA}, we focus on the influence of the Mn-Ga antisite defects on the magnetic properties. From Table~III, comparison of  {\bf D$_{\rm A}$} and {\bf DwA$_{\rm A}$} moments show that the local moments are mostly unchanged while between {\bf D$_{\rm M}$} and {\bf DwA$_{\rm M}$}  the local moments increase slightly in the latter. 
~However, the main reason for the increase in the total moment is the ferromagnetic contribution from Mn$_{\rm Ga}$, although only about 13\%,\cite{Singh12} is sizable because of its large moment (3.38 $\mu_B$). This causes the total moment to increase substantially, {\it e.g.} from 2.22 to 2.85 $\mu_B$/f.u. in the austenite phase and from 1.25 to 2.07 $\mu_B$/f.u. in the martensite phase. Experimentally however, a saturation moment of 1.5 $\mu_B$/f.u. is obtained at 5~K indicating that for such a complicated disordered system with two types of magnetic interactions, SPRKKR  might be overestimating the moment or it is also possible that the structure might not be fully disordered. Moreover, possibility of rotated or tilted Mn magnetic moments  at the interface of the ferromagnetic cluster formed by Mn-Ga antisite defect 
~has been proposed,\cite{Singh12} which has not been considered in the present calculation.


\begin{table*} 
\caption{Spin magnetic moments ($\mu_{B}$/f.u.) of Mn$_{1+x}$Ni$_{2-x}$Ga in the austenite ({\bf D$_{\rm A}$} structure) and martensite phase ({\bf D$_{\rm M}$} structure).} 
\centering
\begin{tabular}{|c|c|c|c|c||c|c|c|c|} \hline
 {composition} & \multicolumn{4} {|c||} {austenite} & \multicolumn{4} {|c|}  {martensite}     \\ \hline
  & Mn$_{\rm Mn}$ & Ni & Mn$_{\rm Ni}$ &  total  & Mn$_{\rm Mn}$ & Ni & Mn$_{\rm Ni}$ &  total      \\ \hline 
  Mn$_2$NiGa ($x$= 1)  & 3.29 & 0.43 & -1.49 &  2.22 & 3.12 & 0.34 & -2.21 &  1.25  \\ \hline         
  Mn$_{1.75}$Ni$_{1.25}$Ga ($x$= 0.75)  & 3.34 & 0.40 & -1.56 &  2.64 & 3.21 & 0.34 & -2.33 &  1.85  \\ \hline                   
  Mn$_{1.5}$Ni$_{1.5}$Ga  ($x$= 0.5) & 3.39 & 0.37 & -1.78 &  3.03 & 3.28 & 0.37 & -2.34 &  2.63   \\ \hline                         
  Mn$_{1.25}$Ni$_{1.75}$Ga  ($x$= 0.25) & 3.45 & 0.36 & -2.08 &  3.54 & 3.35 & 0.39 & -2.36 &  3.41   \\ \hline                         
  Ni$_{2}$MnGa  ($x$= 0) & 3.52 & 0.34 & - &  4.13 & 3.42 & 0.41 & -  &  4.16  \\ \hline                                             
\end{tabular}
\end{table*}

\begin{figure} [htb]
\epsfxsize=90mm
\epsffile{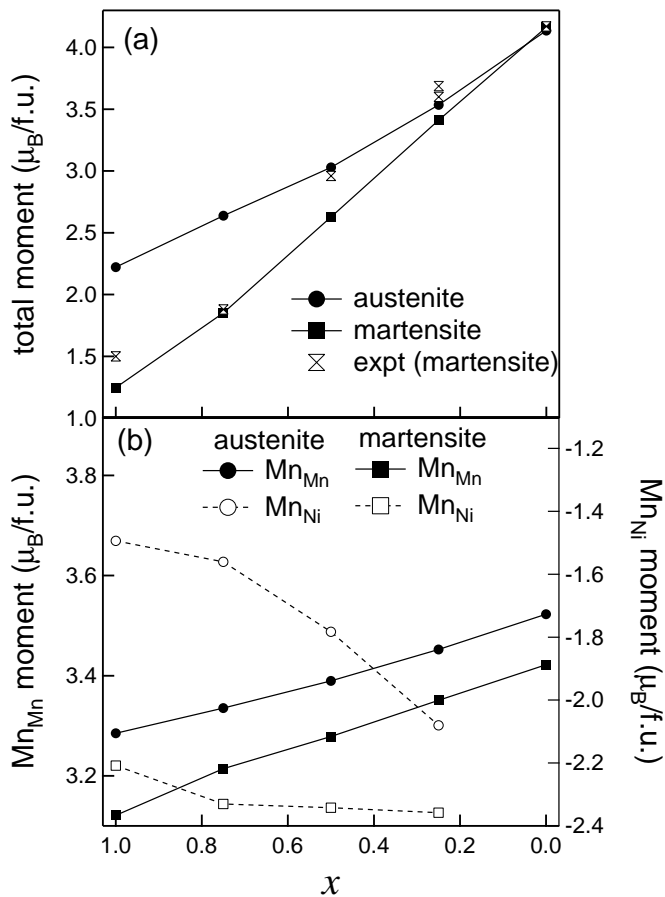}
\caption{(a) The variation of the total magnetic moment of Mn$_{1+x}$Ni$_{2-x}$Ga  as a function of $x$ (1$\geq$$x$$\geq$0) {\it i.e.} decreasing Mn concentration from Mn$_2$NiGa to Ni$_2$MnGa. The  experimental magnetization data\cite{Liu06,Singh12,Banik08} are also indicated by symbols. 
~(b) The variation of the Mn$_{\rm Mn}$ and Mn$_{\rm Ni}$ spin magnetic moments of Mn$_{1+x}$Ni$_{2-x}$Ga  as a function of $x$.} 
\label{fig 1}
\end{figure}


\subsection {The magnetic moments in  Mn$_{1+x}$Ni$_{2-x}$Ga as a function of composition ($x$)}

The total magnetic moment clearly increases with decreasing Mn content  from 2.22~$\mu$B/f.u. in Mn$_2$NiGa ($x$=~1) to 4.13~$\mu$B/f.u. for Ni$_2$MnGa ($x$=~0) in the austenite phase, and similar variation is observed also in the martensite phase (Table V) (see Fig.~1(a)). As $x$ decreases, although the magnetic moment of Ni remains almost unchanged and Mn$_{\rm Mn}$ moment increases while Mn$_{\rm Ni}$ moment decreases  (Fig.~1(b)), the large increase in the total moment is caused by the  decrease in the number of  Mn$_{\rm Ni}$ atoms that reduces the contribution of  Mn$_{\rm Mn}$-Mn$_{\rm Ni}$ antiferromagnetic (AFM) interaction to the total moment. This can be correlated with the decrease in the Mn$_{\rm Ni}$ 3$d$ partial DOS with decreasing $x$, as shown latter in Fig.~5(b). As the composition approaches Ni$_2$MnGa, the Mn$_{\rm Mn}$- Mn$_{\rm Mn}$ ferromagnetic interaction  mediated by Ni dominates and a nearly linear variation of the total moment with $x$ is observed (Fig. 1(a)). 

Having discussed the behavior of the magnetic moments with composition, we turn our attention to what happens across the martensite transition. For Ni$_2$MnGa ($x$=~0), the difference in the total moment between the austenite phase and the martensite phase 
{\it i. e.} $\Delta$$M$ is negative (Fig.~1, Table V). On the contrary, for $x$$\geq$~0.25,  the total moment in the martensite phase is smaller than the austenite phase ($\Delta$$M$$>$~0). This  behavior for Mn excess compositions originates mainly due to the decrease in Mn$_{\rm Ni}$ moments in the martensite phase (Fig.~1(b)). The origin of this effect is related to the martensite phase transformation resulting in the modification of the magnetic exchange interactions due to change in the crystal structure. 
~From Table III, it is evident that the total spin moment is smaller in the martensite phase for both Mn$_2$NiGa-{\bf D} and -{\bf DwA}, the corresponding $\Delta$$M$ being 44\% and 27\%, respectively. Experimentally, about 10\% decrease in magnetization  is observed in Mn$_2$NiGa in both cooling and heating cycles  across the martensite transition at 5 Tesla.\cite{Singh14}  An interesting property of metamagnetic shape memory alloys is inverse magnetocaloric effect in which a magnetic marterial cools down under the application of external magnetic field adiabatically, due to an increase in the isothermal magnetic entropy ($\Delta$S$>$0) of the spin structure.\cite{Ranke09,Manosa13} This effect has been related to the decrease of magnetization in the martensite phase across the martensite transition temperature
~and has been observed in metamagnetic shape memory alloys such as Ni-Mn-Sn and Ni-Mn-In.\cite{Krenke05,Moya07} In contrast, for conventional magnetocaloric effect, magnetization increases in the martensite phase. The sign of $\Delta$M calculated by us could be used to predict  inverse or conventional magnetocaloric behavior, since the change in magnetization across the martensite transition temperature is related to the structural change between the two phases.  It is interesting to note that our calculations for Mn-Ni-Ga systems show that $\Delta$M is positive 
~ in almost the whole range ($x$$\geq$~0.25) of Mn excess compositions predicting that inverse magnetocaloric behavior would be observed. Very recently, this has been corroborated by experiment: both Mn$_2$NiGa and Mn$_{1.75}$Ni$_{1.25}$Ga exhibit inverse magnetocaloric effect, while the conventional magnetocaloric cooling is observed for 
~Ni$_2$MnGa.\cite{Singh14,Devarajan13,Hu00} 

\subsection{Exchange interaction parameter and Curie temperature of Mn$_2$NiGa}
\begin{figure*}[htb]
\epsfxsize=90mm
\epsffile{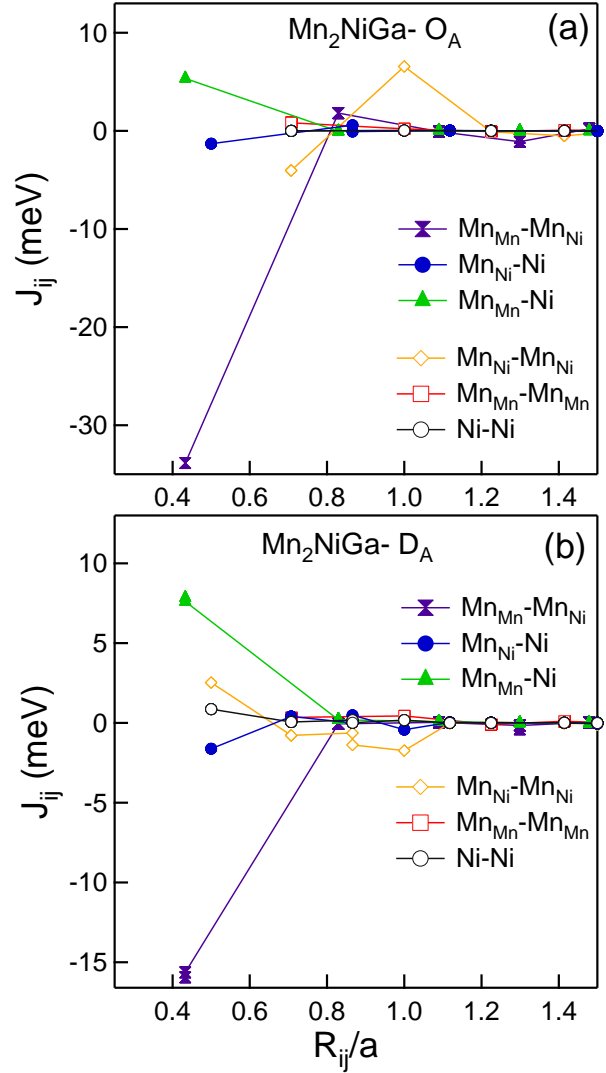}
\caption{Magnetic exchange interaction parameters (J$_{ij}$) calculated using FP-SPRKKR as a function of  distance $R_{ij}$/$a$ ($a$ is the lattice constant) between pairs of atoms $i$ and $j$  for (a) Mn$_2$NiGa-{\bf O$_{\rm A}$} and (b) Mn$_2$NiGa-{\bf D$_{\rm A}$}.}
\label{fig 2}

\end{figure*}

Large Curie temperature of 588~K is one of the major advantages of Mn$_2$NiGa over Ni$_2$MnGa from technological perspective. 
~In this section, we have calculated the exchange interaction parameters of Mn$_2$NiGa based on the real space approach as proposed by Lichtenstein {\it et al.}\cite{Liechtenstein87}  The $J_{ij}$'s  for inter sublattice (Mn$_{\rm Mn}$- Ni, Mn$_{\rm Ni}$- Ni, Mn$_{\rm Mn}$- Mn$_{\rm Ni}$) and the intra sublattice (Mn$_{\rm Mn}$- Mn$_{\rm Mn}$, Mn$_{\rm Ni}$- Mn$_{\rm Ni}$, Ni-Ni) interactions for Mn$_2$NiGa-{\bf O$_A$} are shown in Fig.~2(a).  Mn$_{\rm Mn}$-Ni is clearly the dominant ferromagnetic interaction in the first nearest neighbor (nn) (5.4 meV).  The Mn$_{\rm Mn}$- Mn$_{\rm Mn}$ ferromagnetic interaction, although weaker in the first nn (0.8 meV), extends up to the 4th coordinate shell.  The Ni-Ni interaction is also ferromagnetic in the first nn (0.01 meV). The most dominant antiferromagnetic interaction (-34 meV) in the first nn  between  Mn$_{\rm Mn}$ and Mn$_{\rm Ni}$  is damped in the subsequent co-ordination shells.  Compared to this, the antiferromagnetic Mn$_{\rm Ni}$- Ni interaction is considerably weaker (-1.3 meV) in the first nn.   $J_{ij}$ for  Mn$_{\rm Ni}$- Mn$_{\rm Ni}$ is negative in the first nn (-4.0 meV), which destabilizes the ferrimagnetic alignment of the moments. 
~Based on the calculated $J_{ij}$ values, we determine the $T_C$ of Mn$_2$NiGa-{\bf O$_A$} using mean field approach.\cite{Sasioglu04,Rusz06,Sabiryanov97,Pajda01} The value of $T_C$ turns out to 958~K, which is a 63\% overestimate compared to the experimental $T_C$. 

~Although $T_C$ is generally overestimated in the mean field approach\cite{Sabiryanov97,Pajda01}  and Mn$_2$NiGa with different magnetic interactions is not strictly a Heisenberg system, this large overestimation of $T_C$ needs further investigation because in reality the structure exhibits disorder. Therefore, we have calculated the $J_{ij}$ between different sublattices for the disordered structure (Mn$_2$NiGa-{\bf D$_A$}) as shown in Fig. 2(b) to examine whether this might improve the situation.  Interestingly, we find the main difference with Mn$_2$NiGa-{\bf O$_A$} is that the Mn$_{\rm Ni}$- Mn$_{\rm Ni}$ interaction is  ferromagnetic in the first nn (2.52 meV). Moreover, the magnitude of the first nn Mn$_{\rm Mn}$- Mn$_{\rm Ni}$ antiferromagnetic interaction decreases to -16 meV from -34 meV. The Mn$_{\rm Mn}$- Ni ferromagnetic interaction  increases from 5.3 to 7.8 meV, while the $J_{ij}$ of Mn$_{\rm Ni}$- Ni also changes from -1.3 to -1.6 meV. The value of $T_C$ for Mn$_2$NiGa-{\bf D$_A$} turns out to be 425~K, which is a reasonable estimate that is in better agreement with the experimental value (588~K) compared to Mn$_2$NiGa-{\bf O$_A$}. 
 
 Thus, our FP-SPRKKR calculations show that $T_C$ is overestimated for Mn$_2$NiGa-{\bf O$_A$}, while it is underestimated for Mn$_2$NiGa-{\bf D$_A$}, and similar result was obtained using ASA-SPRKKR.\cite{Sunilcondmat} Previous work on Ni$_{2.25}$Mn$_{0.75}$Ga,\cite{Buchelnikov10} which is a ferromagnet albeit with disorder, also underestimated the $T_C$:  $T_C$ calculated by single site CPA was  200~K, while the experimental value is 352~K. On the other hand,  Ni$_2$MnGa, which is ordered and has one type of magnetic interaction (FM),  the  $T_C$ 
 ~reported  earlier\cite{Sasioglu04,Buchelnikov10,Siewert11} is in much better agreement with the experimental value (376~K).\cite{Banik09}
~Thus, in the case of a disordered system, the averaging character of CPA is a possible reason for the underestimation of $T_C$. Moreover, in Mn$_2$NiGa there are  both FM and AFM  magnetic interactions. 
~Besides, on the experimental front, the value of $T_C$ might depend on the sample heat treatment history as is the case of crystal structure,\cite{Singh10} but such experimental studies do not exist in literature. 







\begin{figure*}[htb]
\epsfxsize=82mm
\epsffile{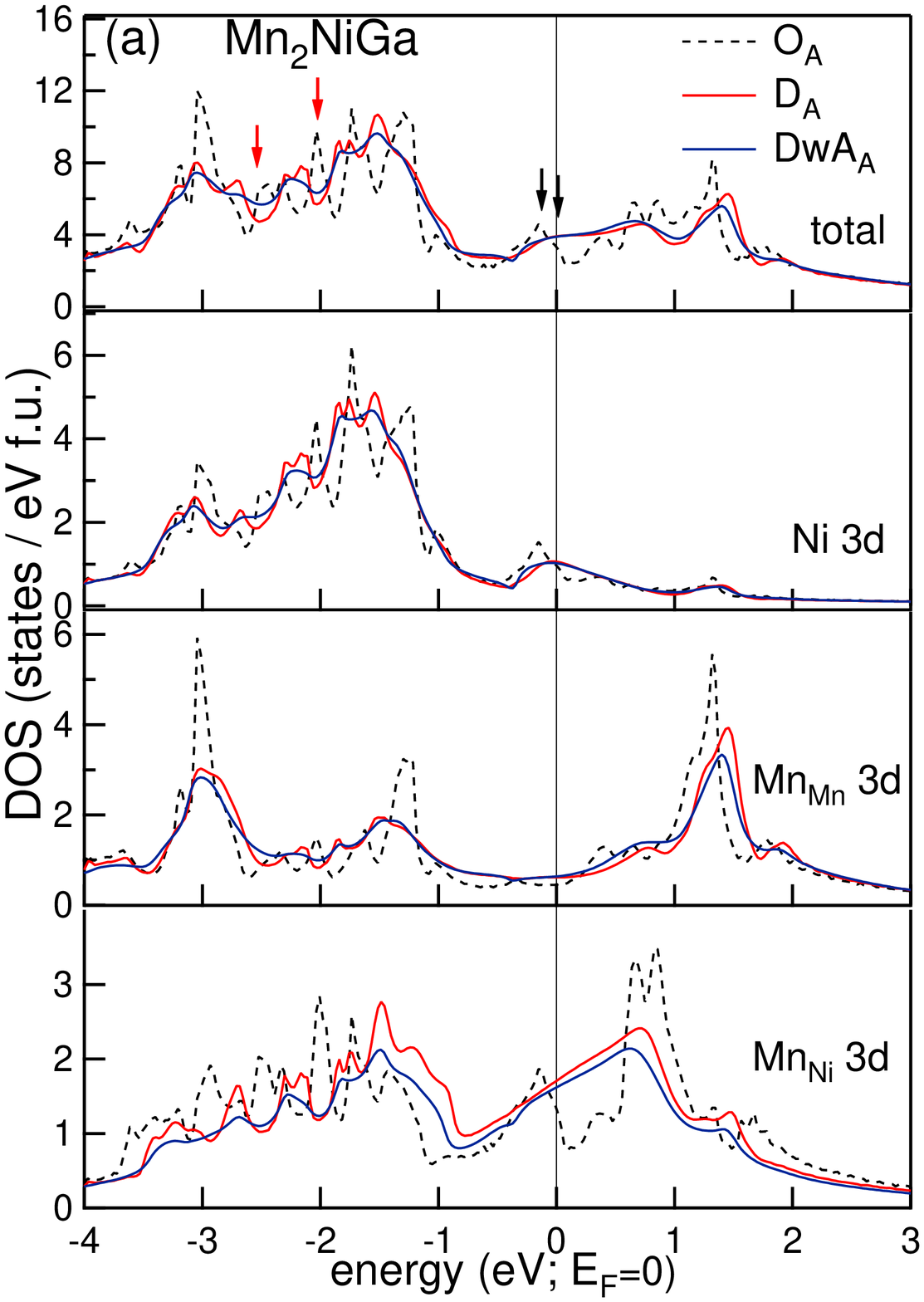}
\epsfxsize=82mm
\epsffile{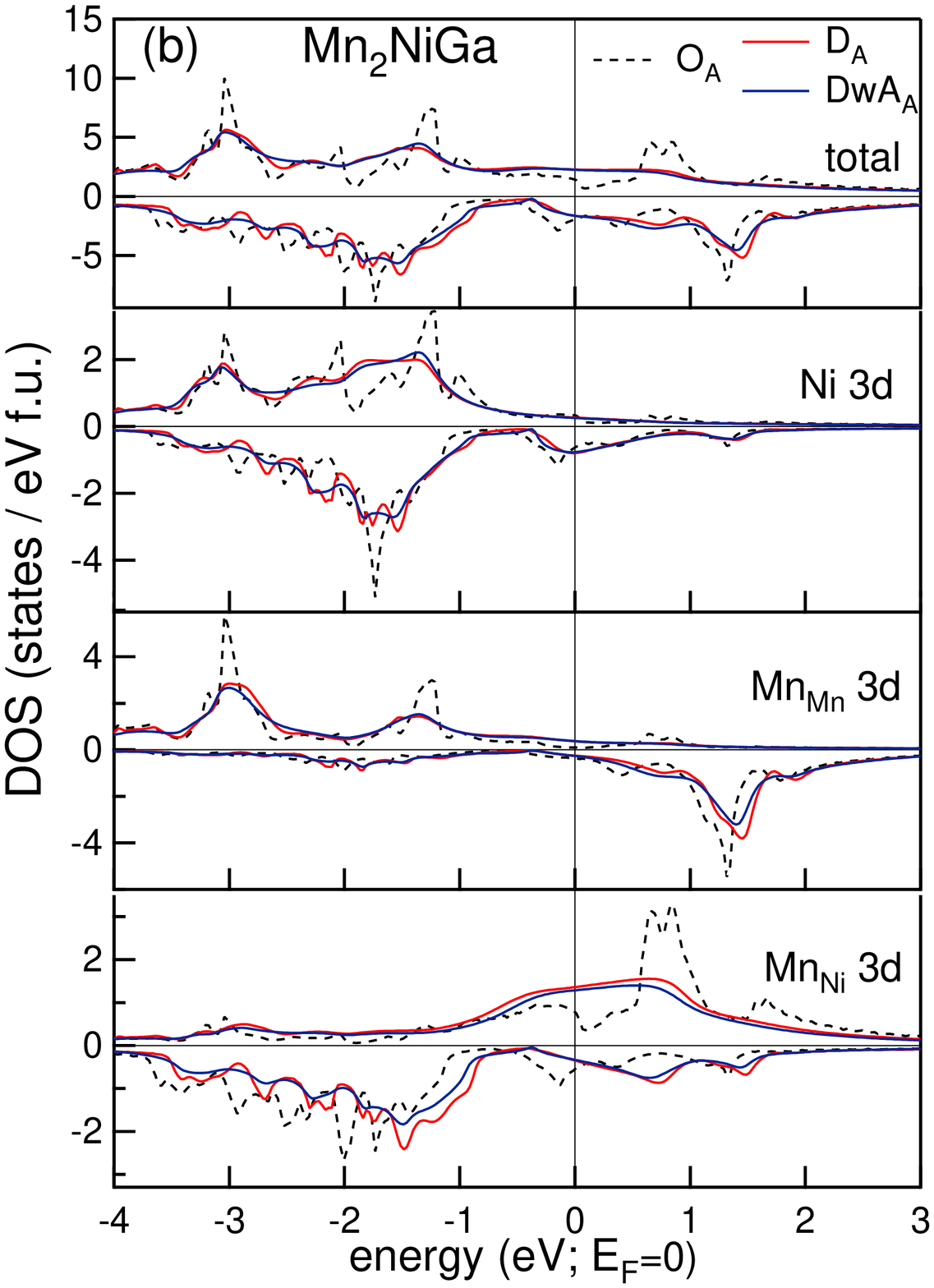}
\caption{Comparison of (a) the total and partial density of states (DOS) and (b) their majority and minority-spin components of Mn$_2$NiGa in ordered ({\bf O$_{\rm A}$}), disordered ({\bf D$_{\rm A}$}) and disordered with antisite defects ({\bf DwA$_{\rm A}$}) structure in the austenite phase.}  
\label{Fig 3}
\end{figure*}

\subsection {Electronic structure of Mn$_2$NiGa: effect of disorder and antisite defects}

The total density of states (DOS) of Mn$_2$NiGa-{\bf O$_{\rm A}$} exhibits a peak close to $E_F$ at -0.15~eV (indicated by arrow) that arises primarily from  hybridization of Ni 3$d$ and Mn$_{\rm Ni}$ 3$d$ minority spin states (Fig.~3). The majority spin Mn$_{\rm Ni}$ 3$d$ also contributes intensity in this region (-0.2~eV) that is manifested as a broad feature intermixed with Ni 3$d$ and Mn$_{\rm Mn}$ 3$d$ majority spin states (Fig.~3(b)). The most intense broad peak centered around -1.5~eV arises mainly from Ni and Mn$_{\rm Ni}$   3$d$ minority spin states, while the majority spin contribution to this peak is due to hybridization of Ni and Mn$_{\rm Mn}$ 3$d$ states. On the other hand, the peak at -3 eV has primarily majority spin character and originates from Ni and Mn$_{\rm Mn}$ 3$d$ majority spin $t_{2g}$ states. The unoccupied DOS is dominated by Mn 3$d$ states; the peak at 0.75~eV is due to Mn$_{\rm Ni}$ 3$d$ majority spin states, while at 1.3~eV the contribution is from Mn$_{\rm Mn}$ minority spin states. In the occupied DOS, the opposite spin peaks dominate: for  Mn$_{\rm Ni}$  the minority spin peak is around -2~eV. From the energy separation of the occupied minority and unoccupied majority spin DOS peaks, we find the exchange splitting of Mn$_{\rm Ni }$ to be 2.75~eV. For Mn$_{\rm Mn}$, the most intense  majority spin peak is around -3~eV, and thus the exchange splitting is estimated to be about 4.4~eV. Antiparallel local moments between Mn$_{\rm Mn}$ and Mn$_{\rm Ni}$ originate from occupancy of primarily  opposite spin states below $E_F$. The exchange splitting energies and the total DOS in Fig.~3 are in agreement with the  
~FPLAPW calculations.\cite{Barman07} 

Interesting modifications are observed in the DOS for Mn$_2$NiGa-{\bf D$_{\rm A}$}.  In the occupied part, the peak at -0.15~eV  in the DOS 
is replaced by a smoothly increasing feature  that is nearly flat at $E_F$ (Fig.~3(a), top panel). The minority spin Mn$_{\rm Ni}$ 3$d$ PDOS peak broadens and shifts closer to $E_F$ and appears at about -1.5~eV. In the unoccupied part, the peak at 0.75~eV that is related to Mn$_{\rm Ni}$ 3$d$ majority spin states is drastically modified: it decreases in intensity and flattens out into a broad plateau over the energy range of -1 to 1~eV merging with the peak at -0.15~eV. These changes imply a sizable decrease in the exchange splitting of Mn$_{\rm Ni}$ to about 2.2~eV.  The Ni 3$d$ minority spin states that hybridize strongly with  Mn$_{\rm Ni}$ 3$d$ minority spin states also broaden and shift by 0.25 eV toward $E_F$ from -1.7 to -1.56~eV.  In contrast, none of the PDOS peaks of  Mn$_{\rm Mn}$ 3$d$ exhibits   any shift, although broadening is clearly observed. This explains why the Mn$_{\rm Ni}$ spin moment changes in presence of disorder, while Mn$_{\rm Mn}$ moment is unaffected (Table III). The exchange splitting of Mn$_{\rm Mn}$ also remains unchanged, while Mn$_{\rm Ni}$  exchange splitting shows large decrease from 2.8 to 2 eV due to disorder. 

Inclusion of Mn-Ga antisite defect in the disordered structure (Mn$_2$NiGa-{\bf DwA$_{\rm A}$}) causes subtle modifications in the DOS. The broadening increases in general: for example, broadening of the -1.5 and -3~eV peaks fills up the valleys around -2 and -2.6~eV (marked by red arrows in Fig.~3(a), top panel).  This is caused by the broadening of the minority spin Ni 3$d$- Mn$_{\rm Ni}$ 3$d$ states. Interestingly,  although the Ni 3$d$- Mn$_{\rm Mn}$ 3$d$ majority spin states have sizable contribution to these peaks, these states hardly  exhibit any broadening. In the unoccupied states also, it is the minority spin states that broaden, see for example the peak at -3~eV originating from Mn$_{\rm Mn}$ 3$d$ states (Fig.~3(b), top and third panel).  Thus, inclusion of antisite defects 
~ marginally broadens the  minority spin PDOS, while majority spin PDOS is hardly affected in both occupied and unoccupied states. 

\begin{figure*}[htb]
\epsfxsize=85mm
\epsffile{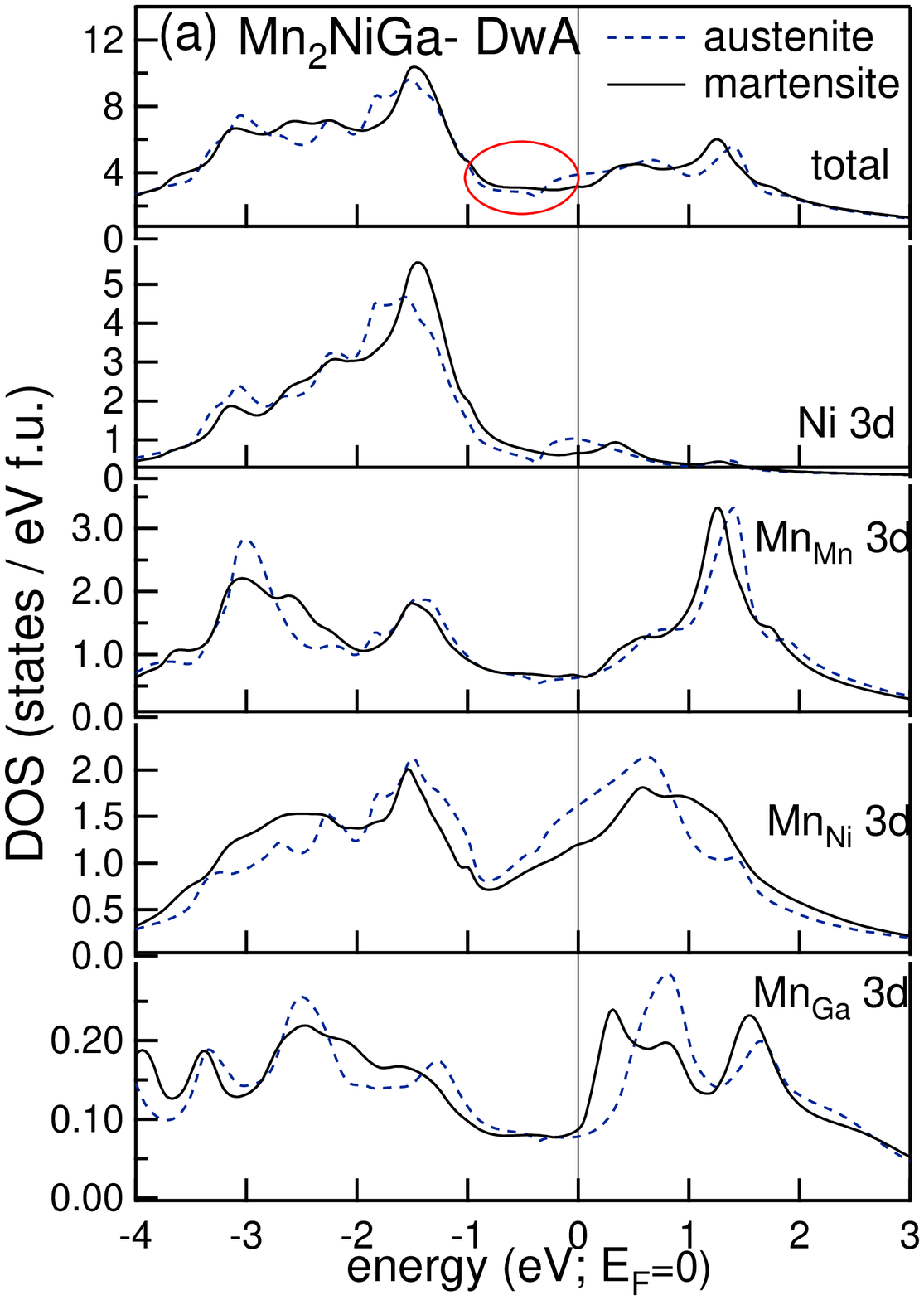}
\epsfxsize=85mm
\epsffile{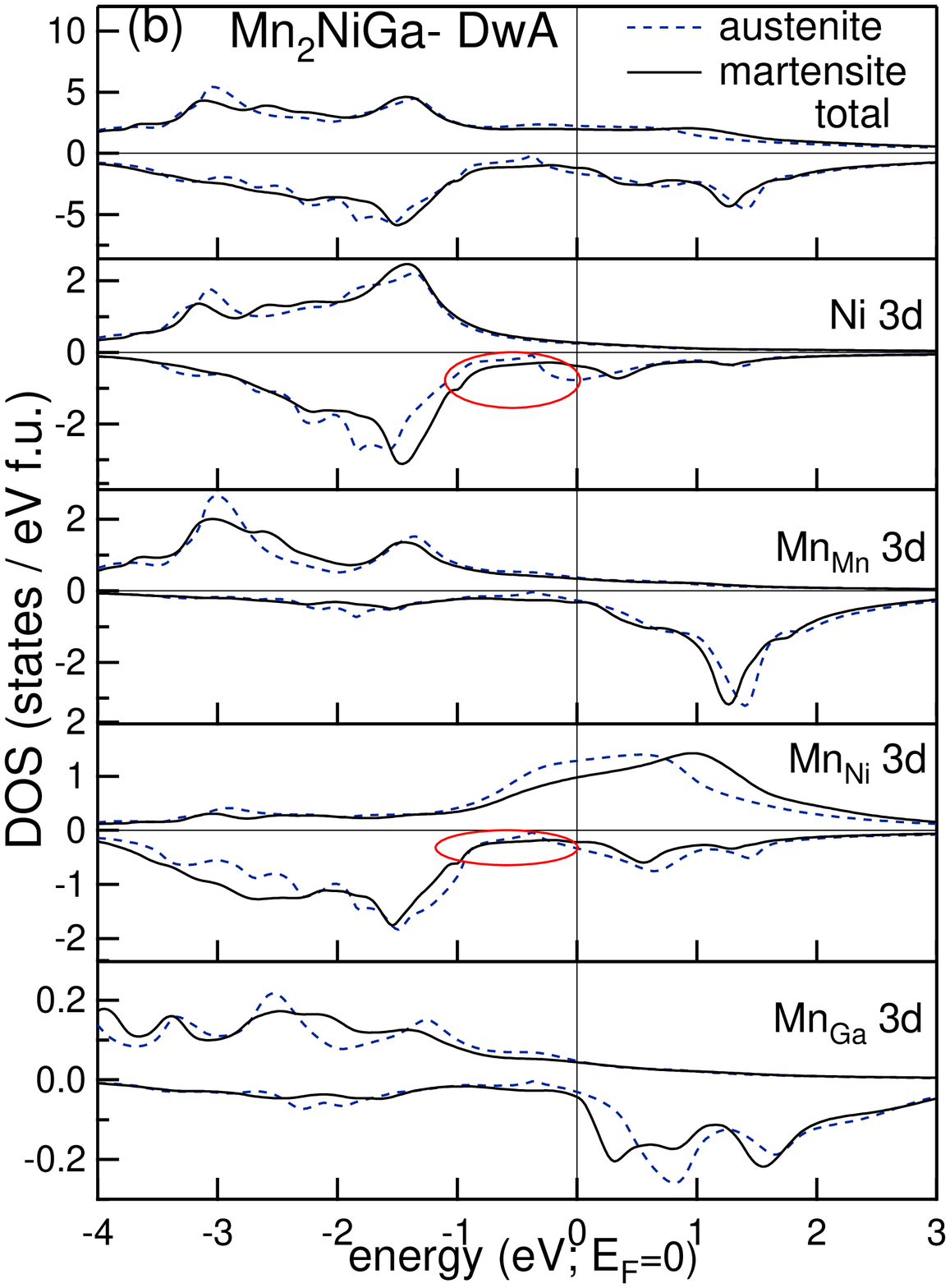}
\caption{Comparison of (a) the total DOS and Ni 3$d$ and Mn 3$d$ PDOS and (b) their majority and minority-spin components between Mn$_2$NiGa-{\bf DwA$_{\rm A}$} (austenite phase) and Mn$_2$NiGa-{\bf DwA$_{\rm M}$} (martensite phase).} 
\label{Fig 4}
\end{figure*}

\subsection {Electronic structure of Mn$_2$NiGa-{\bf DwA} across the martensite transition}

Lattice constant optimization through total energy minimization showed that Mn$_2$NiGa-{\bf O} is stabilized in the martensite phase through a tetragonal distortion of $c/a$= 1.25,\cite{Barman07} which was in agreement with XRD ($c/a$= 1.21). The DOS calculated for the two phases by FPLAPW method showed a peak at -0.1 eV 
~(a corresponding peak at -0.15~eV is also observed in Fig.~3 for {\bf O$_{\rm A}$}).  The shift of this peak  to lower energy in the martensite phase  was related to its  stabilization (see Fig.~3 of Ref.\cite{Barman07}).

Although we use the same lattice constants as in Ref.\cite{Barman07}, the above mentioned peak near $E_F$ is  absent due to disorder  (Fig. 3(a)). So, it is an important question how the states near $E_F$ behave  in the realistic Mn$_2$NiGa-{\bf DwA} structure, since this will provide a clue to the stability of the martensite phase. 
~From the top panel of Fig.~4(a),  we note that the total DOS at E$_F$ in the martensite phase (3.1 states/eV~f.u.) is clearly reduced  in comparison to the austenite phase (4.0 states/eV f.u.). In fact, a suppression of the DOS in the martensite phase is observed between -0.35~eV to $E_F$, while, in contrast, the DOS between  -1.0 and -0.4~eV is enhanced (see the encircled region in 
~Fig.~4(a), top panel). 
~Thus, there is an unambiguous evidence of transfer of electron states from higher to lower energies  that would stabilize the martensite phase.  Such transfer of electron states to lower energy is also observed for Mn$_2$NiGa-{\bf D} (Fig.~5(a), top panel). In fact, the 
~total DOS for both  Mn$_2$NiGa-{\bf DwA} and -{\bf D} structures not only shows the transfer of states in the energy range of -1 eV to $E_F$, but also in the -2.8 to -1.5~eV region. 

~From the PDOS (Fig.~4(b)), we find that the  Ni 3$d$- Mn$_{\rm Ni}$ 3$d$ minority spin states are primarily responsible for such redistribution of the electron states, as shown by the red ellipses in Fig.~4(b).  This can be related to the tetragonal distortion ($c/a$= 1.25) in the martensite phase because the Ni-Mn$_{\rm Ni}$  distance decreases to 2.70\AA~ from 2.93\AA~ in the austenite phase resulting in stronger hybridization. 

The majority spin Mn$_{\rm Ga}$ 3$d$ states  appear around -1.3 and -2.5 eV. The minority spin PDOS of Mn$_{\rm Ga}$ above $E_F$  shows a substantial shift toward $E_F$ in the martensite phase, thus causing an increase of states at $E_F$. Although the behavior of Mn$_{\rm Ga}$ 3$d$ PDOS is opposite to the behavior of Ni 3$d$- Mn$_{\rm Ni}$ 3$d$ states, its contribution is an order of magnitude small (Fig.~4(b)). The exchange splitting of Mn$_{\rm Ga}$ 3$d$ states is about 3.3~eV, which is somewhat smaller than Mn$_{\rm Mn}$ 3$d$ states. 

The occupied DOS is dominated by the majority  spin PDOS, while the unoccupied DOS is dominated by the minority spin PDOS for both Mn$_{\rm Ga}$ and Mn$_{\rm Mn}$ 3$d$ states. This is the reason that their moments are positive and parallel (Table IV). In contrast, the occupied (unoccupied) DOS is dominated by the majority  spin PDOS, while the unoccupied (occupied) DOS is dominated by the minority spin PDOS for Mn$_{\rm Mn}$ (Mn$_{\rm Ni}$) 3$d$ states leading to antiparallel orientation.

\subsection {The electronic structure of Mn$_{1+x}$Ni$_{2-x}$Ga as a function of composition}
\begin{figure*}[htb]
\epsfxsize=82mm
\epsffile{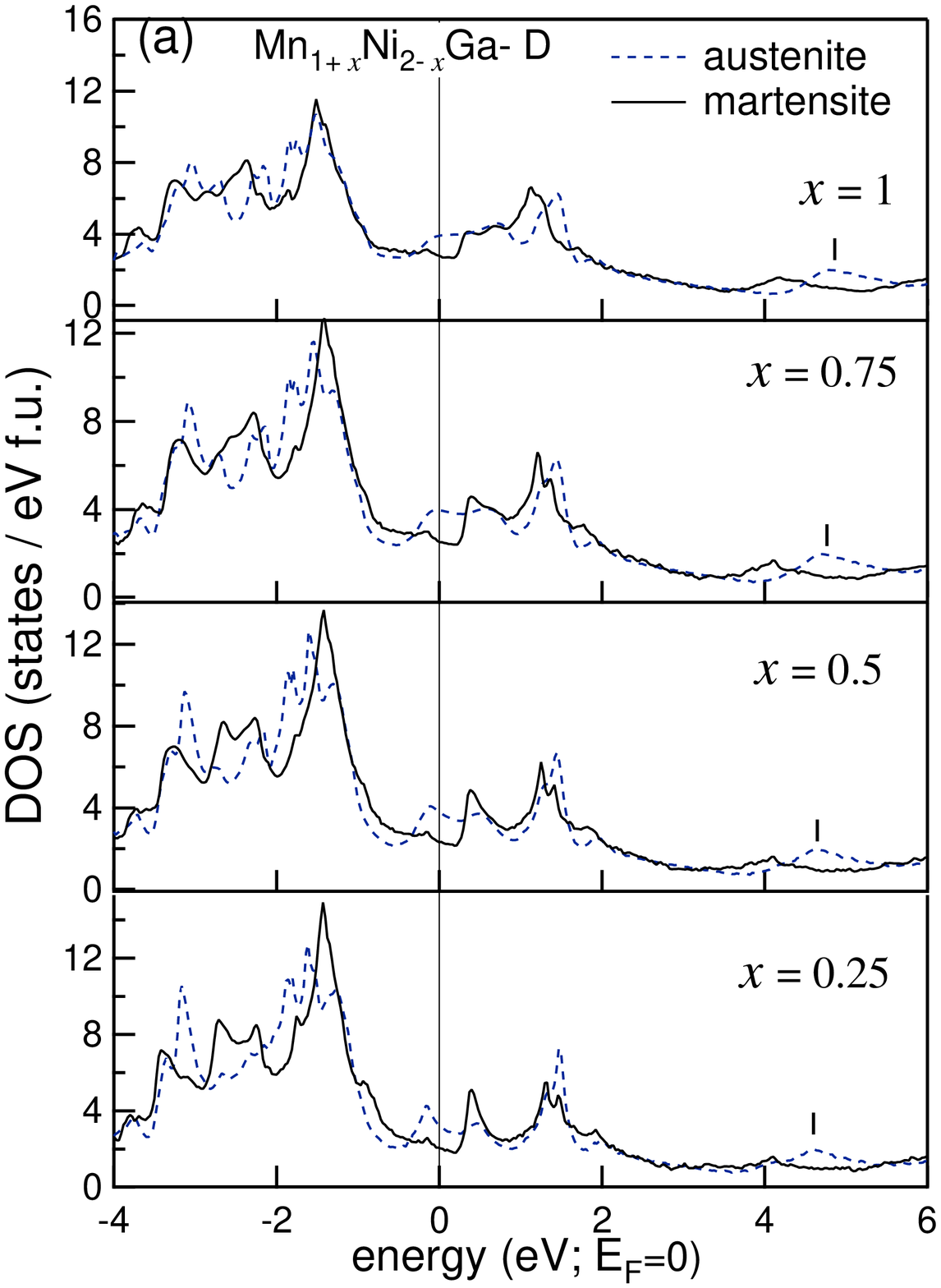}
\epsfxsize=82mm
\epsffile{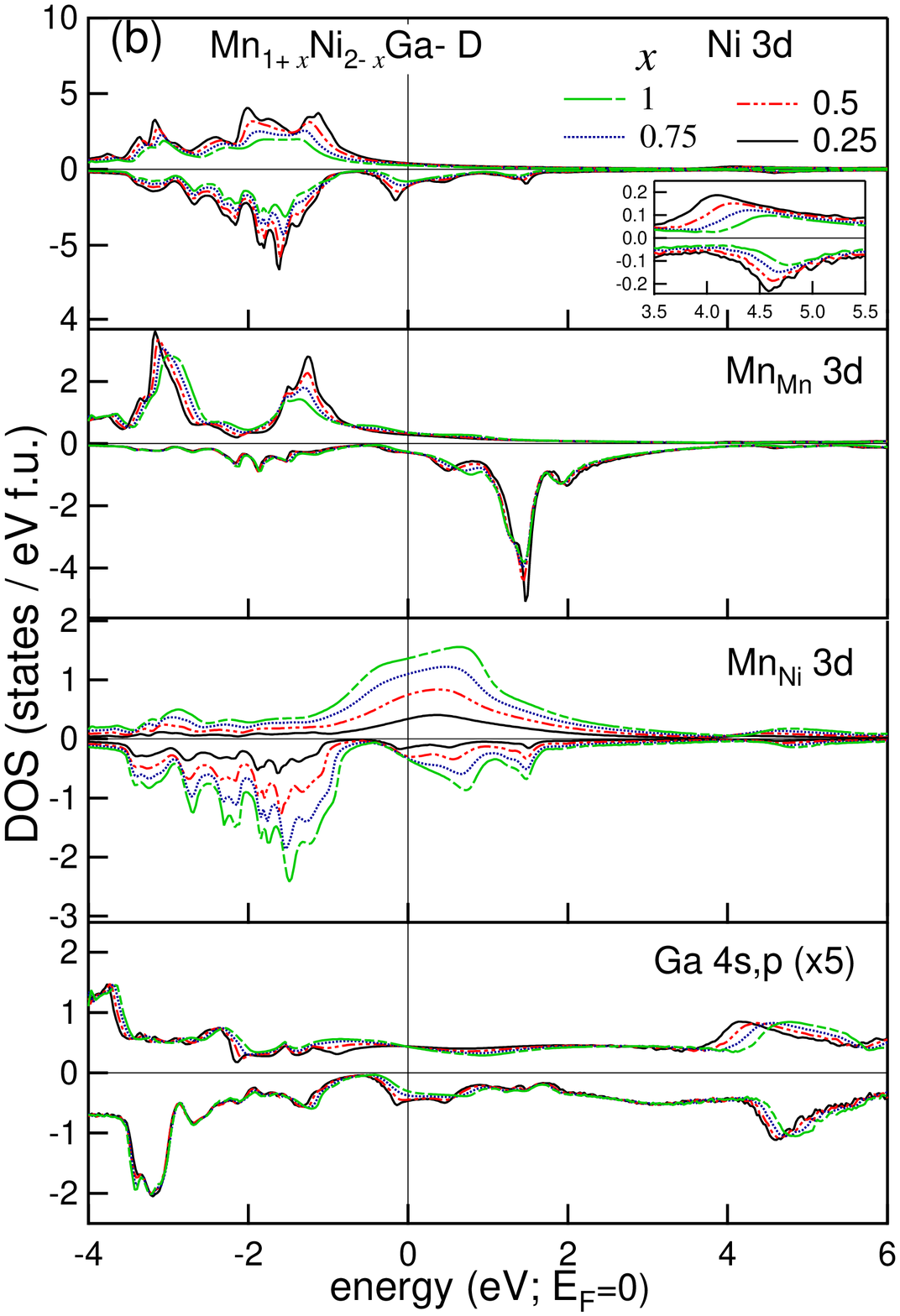}
\caption{Comparison of (a) the total DOS in the austenite and martensite phases; the majority and minority spin components of Ni 3$d$, Mn 3$d$ and Ga 4$s,\,p$ PDOS in the (b) austenite phase of Mn$_{1+x}$Ni$_{2-x}$Ga, as a function of $x$ (1$\geq$$x$$\geq$0).}  
\label{Fig 5}
\end{figure*} 

In order to understand the role of disorder on the electronic structure of Mn$_{1+x}$Ni$_{2+x}$Ga, we have calculated the total and partial density of states (PDOS) for the series 1$\geq$$x$$\geq$0 (Figs.~5). Here, $x$ quantifies Mn$_{\rm Ni}$ in the formula unit (f.u.) and thus as $x$ decreases, the  disorder also decreases. 
Since martensite transition is observed for the whole series,\cite{Liu05,Banik09}  the austenite and martensite total DOS are compared to understand the origin of the stability in the latter (Fig. 5(a)). The Mn-Ga antisite defects are not considered since this has not been experimentally studied for the whole series.    Transfer of electron states to lower energies in the range -0.8~eV to $E_F$ similar to that of Mn$_2$NiGa is noticed for all the compositions, indicating similar origin of the martensite transition in this series.

An interesting change in the DOS near $E_F$ as disorder {\it i.e.} $x$ decreases is the appearance of a  sharp peak below $E_F$ in  the austenite phase at -0.16~eV for $x$= 0.75-0.25. This peak is   primarily related to Ni 3$d$ e$_g$ minority spin states with sizable admixture of Mn$_{\rm Ni}$ 3$d$ states. 
~This peak also broadens out as disorder increases and is hardly present in Mn$_2$NiGa (Fig. 5(a)). 
In Fig.~3(a) top panel, the peak observed at -0.15 eV in Mn$_2$NiGa-{\bf O$_{\rm A}$}  disappears in Mn$_2$NiGa-{\bf D$_{\rm A}$}. Thus, it is evident that this  peak close to $E_F$ arising from Ni 3$d$ e$_g$ minority spin states is strongly affected by disorder. This peak near $E_F$ has not been observed in the photoemission spectrum of Mn$_2$NiGa,\cite{Barman07} and it is evident from the present work that it's absence is related to disorder.
The energy separation between the occupied majority and the unoccupied minority 
peaks of Mn$_{\rm Mn}$ PDOS exhibits a small increase in the exchange splitting energy from about 4.5 eV ($x$= 1) to  4.7 eV ($x$= 0.25). On the other hand, the Mn$_{\rm Ni}$ PDOS shows a  decrease in the exchange splitting  from  2.3 eV ($x$= 1) to 2.0 eV ($x$= 0.25).  

\section {Conclusion}
Using full potential spin-polarized scalar relativistic Korringa-Kohn-Rostocker (SPRKKR) method, we have studied the spin  moments, exchange parameters, Curie temperature and the spin polarized DOS and partial DOS of Mn$_2$NiGa in structures such as ordered ({\bf O}), disordered ({\bf D}) and disordered with antisite defect ({\bf DwA}). Moreover, the series Mn$_{1+x}$Ni$_{2-x}$Ga has been studied in the {\bf D} structure. For Mn$_2$NiGa, the
total spin moment increases due to disorder because of the decrease in the magnitude of the antiparallel Mn$_{\rm Ni}$ moment. The presence of Mn-Ga antisite defects induces ferromagnetic interaction between Mn$_{\rm Mn}$ and Mn$_{\rm Ga}$ atoms that enhances the total moment. 
~For Mn$_2$NiGa, total spin moment decreases in the martensite phase for both the disordered as well as antisite defect structures compared to the austenite phase. 
~The exchange parameters show interesting difference between  ordered and disordered  Mn$_2$NiGa. A reasonable estimate of $T_C$ (425 K) compared to the experimental value (588 K) is obtained for the disordered structure (Mn$_2$NiGa-{\bf D$_A$}).  
~Disorder influences the electronic structure of Mn$_2$NiGa through overall broadening of the PDOS and a decrease in the exchange splitting. Inclusion of antisite defects in the calculation marginally broadens the minority spin PDOS, while the majority spin PDOS is hardly affected in both the occupied and unoccupied states.  

For Mn$_{1+x}$Ni$_{2-x}$Ga, as $x$ decreases, 
~an increase in the total moment  
~ is caused by the decrease in the number of  Mn$_{\rm Ni}$ atoms that reduces the contribution of  Mn$_{\rm Mn}$-Mn$_{\rm Ni}$ antiferromagnetic interaction. While for Ni$_2$MnGa, the  total spin moment in the martensite phase is larger than the austenite phase,  for $x$$\geq$~0.25 this is reversed {\it i.e.} the moment in the martensite phase is smaller. This indicates possible occurrence of inverse magnetocaloric behavior for $x$$\geq$~0.25.  Mn$_{\rm Mn}$ PDOS exhibits an increase in the exchange splitting energy as $x$ decreases, 
~while the Mn$_{\rm Ni}$ PDOS shows a decrease in the exchange splitting. 
~A redistribution of Ni 3$d$- Mn$_{\rm Ni}$ 3$d$ minority spin  electron states near $E_F$ is primarily responsible for the stability of the martensite phase in Mn-Ni-Ga. 

\section{Acknowledgment}

H. Ebert, M. Meinert,  J. Min$\acute{a}$r, S. Mankovsky, B. Sanyal, and S. Singh are thanked for useful discussions.  S.W.D. thanks Council of Scientific and Industrial Research, New Delhi for research fellowship. The support from the scientific computing group of the Computer Centre, RRCAT is greatfully acknowledged.



\end{document}